# From Steady-State to Ultrafast: Resonance Raman Approaches for Biological Samples

Juan J. Romero, Bruno Robert, Manuel J. Llansola-Portoles*

*Université Paris-Saclay, CEA, CNRS, Institute for Integrative Biology of the Cell (I2BC), 91190, Gif-sur-Yvette, France.*

**Abstract**

Vibrational spectroscopy reports on molecular structure with chemical-bond specificity, but in biological systems the vibrational signals of a target chromophore are typically buried under contributions from the surrounding matrix. Resonance Raman (RR) spectroscopy addresses this problem by matching the excitation wavelength to an electronic transition of the chromophore of interest, which increases Raman cross sections by up to six orders of magnitude and restricts the enhanced modes to those coupled to the resonant electronic state. This chapter introduces the physical basis of resonance enhancement and shows how RR isolates chromophore-specific vibrational markers in complex biological systems. We then extend the same principle into the time domain with femtosecond stimulated resonance Raman spectroscopy (FSRRS). In FSRRS, the Raman pump is tuned across the visible range and can be placed in resonance with the transient absorption of a chosen excited-state species, which makes the Raman pump wavelength an additional experimental variable. Sampling this variable across the excited-state absorption manifold provides a selection criterion that separates coexisting transient species sharing a common vibrational window.

*Corresponding author: manuel.llansola@cnrs.fr

INDEX

# 1. Introduction

This chapter is addressed to biologists and biophysicists who work with chromophore-containing systems and need to extract structural and environmental information from vibrational spectra. Resonance Raman (RR) spectroscopy provides chemical selectivity in the ground state; while femtosecond stimulated resonance Raman spectroscopy (FSRRS) extends that selectivity into the femtosecond time domain and adds a further analytical dimension, the Raman pump wavelength, which can be tuned across the transient absorption manifold of a photoexcited sample. We first introduce Raman scattering concepts needed to interpret spectra, and in particular explain why resonance conditions favour both sensitivity and selectivity. We then show, through selected examples, how RR isolates chromophore signatures in pigment-protein complexes and related biological contexts where many species coexist. We then describe FSRRS and the role of the Raman pump wavelength as a selection variable, and illustrate the approach with molecules of biological interest like carotenoids. The chapter closes with a practical guide to the experimental constraints that appear across biological applications of both techniques. FSRRS is a recent approach, and the examples available on macromolecular systems remain few, so a review of the subject is more adapted than a detailed protocol.

# 2. Raman basics for non-specialists

## 2.1. Raman Scattering

Vibrational spectroscopy gives access to the energy of vibrational levels of molecules, which are extremely sensitive to their structural and electronic configuration. One of the main approaches for vibrational spectroscopy is Raman scattering. This phenomenon involves a change in the frequency of light when it is scattered, reflecting energy exchange between the incoming photon and the scattering molecule. Classically, the Raman effect can be described as an interaction between the oscillating electric field of the incident light, which disturbs the electronic charge distribution in the molecule, and the intrinsic oscillations (normal modes) of the illuminated molecule. The product of these two oscillating functions can be approximated by a cosine product, which can be written in the form of three cosine functions possessing: the frequency of the incident light; the sum of the frequency of the incident light and that of each vibrational mode which influences the molecular polarisability; and the difference between these frequencies (Figure 1). This simplified view predicts the appearance in the scattered light of new frequencies, shifted from the energy of the incident light by the energy of the vibrational modes (*1*).

When reasoning in terms of quantised energy levels, the Raman effect corresponds to an ultrafast interaction leading the molecule to a virtual state, during which it may either gain or lose the energy of a vibrational quantum (Stokes or anti-Stokes scattering, respectively; Figure 1). Raman spectroscopy can thus be used as an analytical method for determining the chemical structure of molecules, as well as their conformation and the intermolecular interactions in which they are involved. Also, considering the underlying selection rules, one practical advantage of Raman spectroscopy over infrared vibrational absorption is that water contributes little to the Raman signal of biological molecules.

Typically, a Raman setup requires a low-power continuous-wave laser system and a spectrally resolved detector to capture the Raman spectrum, which consists of distinct sharp peaks, appearing at wavelengths shifted from the exciting laser wavelength by the energy of the sample's molecular vibrations. Moreover, the use of lasers as exciting sources drastically reduces the sample volume (typically a few microlitres), determined by the size of the focused laser beam. The units used for describing Raman spectrum correspond to the shift between the energy of the photons provided by the exciting laser beam and those emitted by the scattering molecule, expressed in wavenumbers ($cm^{-1}$) for convenience. At room temperature, high-energy vibrational modes are poorly populated, so Raman spectroscopy is usually performed under Stokes conditions, where the analysed scattered light lies at lower energy than the excitation laser beam. Anti-Stokes signals are weaker under ambient conditions but can be informative as a probe of local heating or vibrational population in time-resolved experiments.

The Raman effect is a low-probability process; the signal is weak and easily contaminated by spurious fluorescence. In chromophore-rich biological systems, fluorescence backgrounds from pigments, cofactors, or impurities can be orders of magnitude stronger than the Raman signal, masking vibrational features entirely. This imbalance often makes conventional Raman impractical unless concentrations are sufficiently high or acquisition times are long enough, conditions that might lead to photodamage. One possible option to enhance Raman signals is the use of non-linear optical effects, where a secondary laser interacts with the sample exchanging the extra input energy between optical fields and molecular vibrations, as in the case of *Coherent Anti Stokes Raman Scattering* (CARS) or *Stimulated Raman Scattering* (SRS) (*2*). Particularly, the latter constitute the cornerstone for advanced time resolved Raman techniques. SRS is also free of the non-resonant background that complicates CARS spectra, which simplifies the analysis (*3*). Another possibility for signal enhancement but also for chemical selectivity is the resonance effect, which is described in the next section.

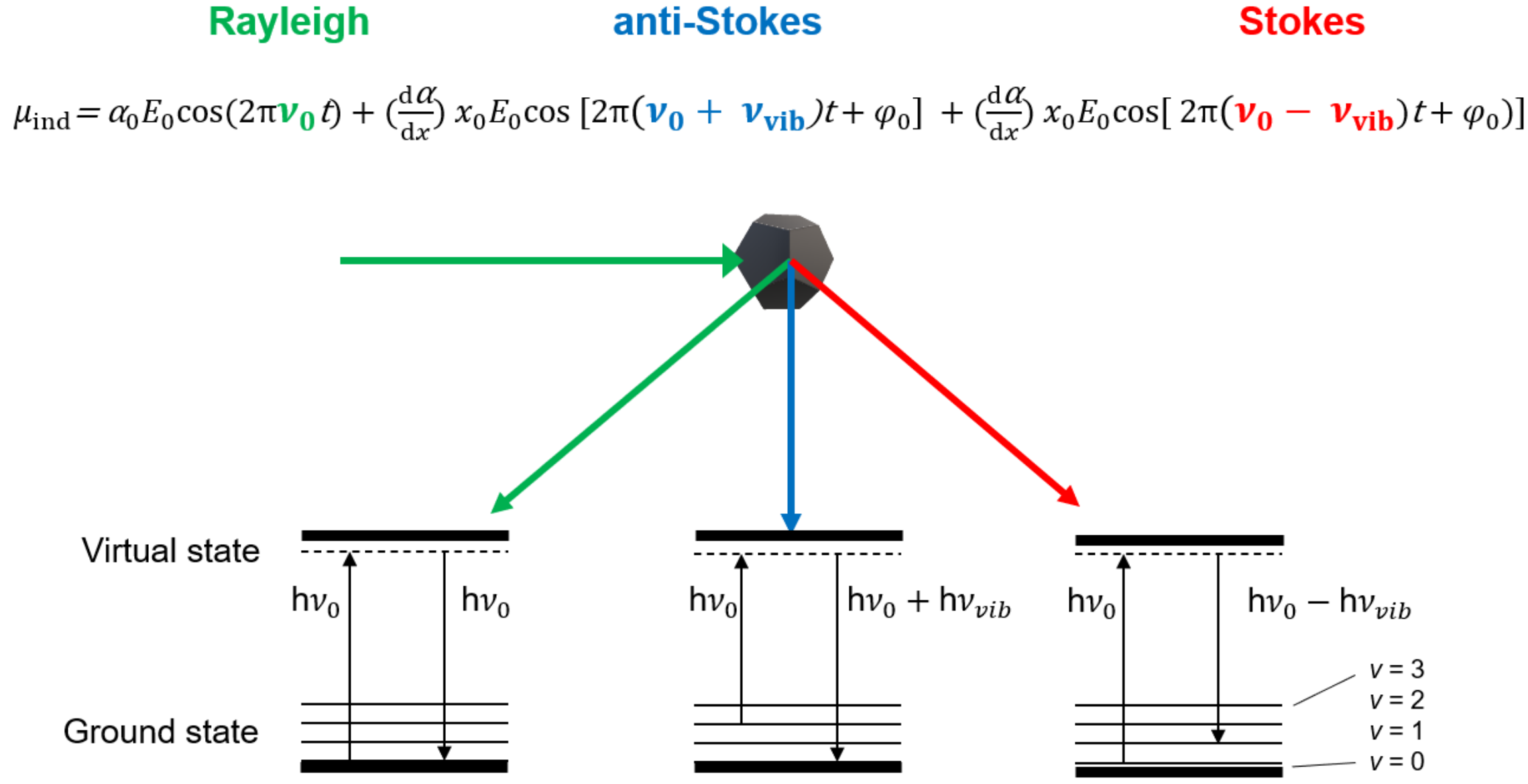

**Figure 1| Principle of Raman scattering.** (Top) Classical description: the induced dipole moment $\mu_{ind}$ contains three frequency components corresponding to Rayleigh ($\nu_0$), anti-Stokes ($\nu_0 + \nu_{vib}$), and Stokes ($\nu_0 - \nu_{vib}$) scattering. (Bottom) Energy-level description of the same three processes. In each case, the incident photon ($h\nu_0$) promotes the molecule to a virtual state (dashed lines), from which it returns to the ground electronic state (solid lines). In Rayleigh scattering, the molecule returns to the same vibrational level and no energy is exchanged. In anti-Stokes Raman scattering, the molecule returns to a lower vibrational level (v = 0), releasing a photon of energy $h\nu_0 + h\nu_{vib}$. In Stokes Raman scattering, the molecule returns to a higher vibrational level (v = 1), releasing a photon of energy $h\nu_0 - h\nu_{vib}$. The vibrational levels of the ground electronic state are indicated on the right (v = 0 to v = 3).

## 2.2. Resonance Raman: selectivity and enhancement

In classical Raman spectroscopy, the signal intensity varies as $\nu_0^4$, where $\nu_0$ is the excitation frequency. When $\nu_0$ matches the frequency of an electronic transition of the irradiated molecule, however, a subset of Raman-active modes is enhanced by up to six orders of magnitude. This is the resonance Raman (RR) effect. A detailed account of the physical origin of this enhancement is beyond the scope of this chapter; the reader is directed to references (*4*, *5*) for a full treatment. The only vibrational modes enhanced under resonance conditions are those involving nuclear motions that correspond to distortions experienced by the molecule during the transition between the ground and excited states used to induce the resonance (*6*). Three consequences of this resonance condition are directly relevant to biological applications.

First, the increase in scattering cross section (gains of $10^4$ to $10^6$) means that a chromophore present at nanomolar concentration in a complex biological sample can yield vibrational signals comparable in intensity to those of the millimolar-concentration range in the same matrix. It is thus possible to selectively observe a chromophore in a complex medium, provided it possesses an absorption transition whose energy matches that of the incoming photons. This makes it possible to study the conformation and interactions of chromophores within proteins, even when those proteins remain embedded in a biological membrane, are incompletely purified, or are probed directly *in vivo*.

Second, the enhancement is mode-selective. Under resonance conditions, only the vibrational modes involving nuclear motions coupled to the resonant electronic transition contribute significantly to the spectrum. This intra-mode selection may appear to be a limitation: if a chemical group is not directly involved in the electronic transition (for instance, a non-conjugated carbonyl), resonance Raman will yield no information about it. In practice, the functional part of most biological chromophores consists precisely of those atoms conjugated with the electronic transition. Resonance Raman therefore reports selectively on the biologically relevant vibrational degrees of freedom of these molecules. In addition, the positions and intensities of the resonance-active bands carry complementary information: band positions report on the vibrational structure of the electronic states involved, while band intensities report on the coupling of each mode with the electronic transition.

Third, when more than one chromophore is present in the sample, each with a distinct electronic absorption, the resonance condition acts as a species-selective filter. Tuning the excitation wavelength onto the absorption of one chromophore and away from others preferentially enhances the vibrational spectrum of the targeted species. In photosynthetic systems, for example, excitation in the carotenoid absorption region (circa 430 to 560 nm, depending on the carotenoid) selectively enhances carotenoid modes, while excitation in the chlorophyll Soret (circa 400 to 420 nm) region enhances chlorophyll contributions. Because resonance selectivity is dependent on the absorption, excitation wavelength is the dominant experimental control parameter.

## 3. Resonance Raman in biological samples

In biological samples, the question is usually not "what molecules are present?" but "what is the conformation, interaction pattern, or microenvironment of a specific chromophore in its functional context?". The examples below illustrate how RR answers that question, in each case by using excitation wavelength as the primary selectivity tool and a small set of band markers as readouts. In every case, the resonance condition is what makes the measurement feasible: without it, the chromophore signal would be buried under contributions from protein, lipid, and solvent.

### 3.1. The resonance principle: species selectivity in a multi-pigment protein

The principle of excitation-wavelength selectivity is illustrated directly by the light-harvesting complex II (LHCII), the major antenna complex of plants and green algae. Each LHCII monomer binds 8 chlorophyll a, 6 chlorophyll b, and 4 carotenoid molecules within a single protein scaffold (Figure 2A) (*7*). In the absorption spectrum of the LHCII trimer, the contributions of the three pigment classes overlap substantially: carotenoids absorb between approximately 430 and 510 nm, Chl *b* near 470 and 650 nm, and Chl *a* near 435 and 680 nm (Figure 2B). From a conventional spectroscopic standpoint, this congestion makes it difficult to isolate the vibrational signature of any individual pigment class. Resonance Raman is an excellent approach to study this kind of systems. Tuning the excitation wavelength to the carotenoid absorption region selectively enhances carotenoid vibrational modes, yielding a spectrum dominated by bands characteristic of carotenoids, with no significant contribution from chlorophyll (Figure 2C, bottom right). Shifting the excitation to the Chl a Soret band instead produces a spectrum in which chlorophyll carbonyl and macrocycle modes dominate (Figure 2C, top right). The same sample, unchanged, yields two chemically distinct vibrational reports depending solely on the excitation wavelength. This is the practical expression of the resonance selectivity. As the following sections show, this selectivity is not limited to distinguishing between chemically different pigment classes. Within a single class, and even among molecules of identical chemical structure, excitation wavelength can be used to tune the Raman response toward subpopulations that differ only in their protein environment or conformational state. Exploiting this resolution requires, however, a precise prior knowledge of the vibrational markers of each pigment type in isolation. The two sections below establish that foundation for carotenoids and chlorophylls before the protein case studies are presented.

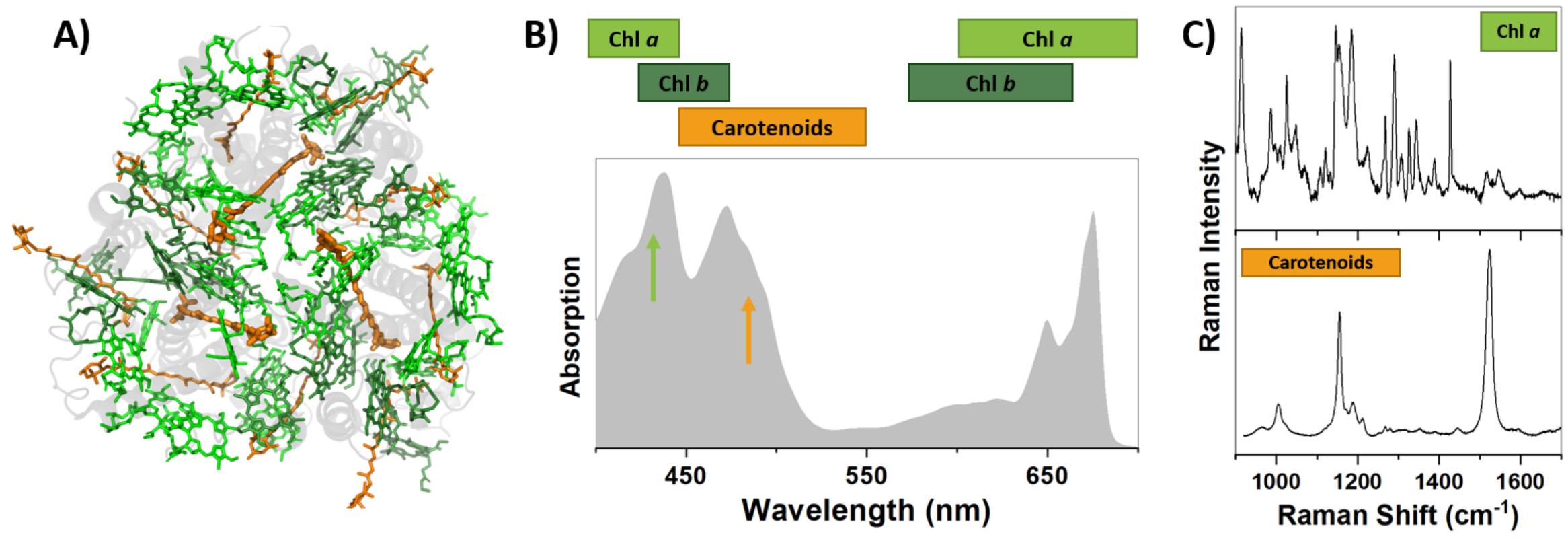

**Figure 2| (*A*)** Crystal structure of the LHCII trimer, containing 14 chlorophylls (Chl a, green; Chl b, dark green) and 4 carotenoids (orange) per monomer (*7*). **(*B*)** Absorption spectrum of LHCII, with the absorption regions of

each pigment class indicated. The coloured arrows mark the position for resonance Raman excitation. ***(C)*** Resonance Raman spectra obtained by tuning the excitation wavelength to the Chl a Soret band (top) or to the carotenoid absorption band (bottom), demonstrating species-selective enhancement from the same sample.

### 3.1. Isolated pigments: Carotenoids and Chlorophylls

Carotenoids are a large family of natural molecules, with more than 1100 members characterised to date, and are particularly favourable RR targets because the strong $S_0$ to $S_2$ electronic transition in the visible region provides efficient resonance enhancement. Carotenoids display the highest resonance Raman cross sections among natural biomolecules, and their spectra are organised into four well-established band regions ($\nu_1$ to $\nu_4$) that together provide a compact structural readout of the conjugated polyene backbone (Figure 2C). The first resonance Raman spectra of an all-trans carotenoid were recorded in 1970 (*8*), and the origin of the four band groups was subsequently assessed by comparison with polyene spectra (*9*). Normal coordinate calculations followed in the 1980s (*10*), and precise spectral prediction is now possible using density functional theory combined with QM/MM or ab initio molecular dynamics approaches, to the point that spectra of phenyl-substituted carotenoid derivatives were predicted before they were measured (*11-15*).

The $\nu_1$ band, above 1500 cm$^{-1}$, is dominated by C=C stretching of the conjugated chain and is sensitive to effective conjugation length and configuration. Because the C=C stretching coordinate is strongly coupled to the $S_0$ to $S_2$ transition, this mode is among the most intensely enhanced under resonance conditions. The $\nu_2$ region, around 1160 cm$^{-1}$, contains C–C stretching coupled to C–H in-plane bending and is used to detect cis content, with specific marker contributions reported for common cis configurations. The $\nu_3$ region, near 1000 cm$^{-1}$, arises largely from methyl rocking and can report on terminal ring configuration where relevant. The $\nu_4$ region, around 960 cm$^{-1}$, includes out-of-plane wagging and torsional motions that gain intensity when the polyene backbone is distorted out of planarity, making $\nu_4$ a probe of backbone twisting and protein-induced distortion. This sensitivity is itself a resonance effect: distortion lowers local symmetry, couples additional coordinates to the electronic transition, and increases the Raman cross section of modes that would be weak in a planar molecule.

Chlorophyll (Chl) and bacteriochlorophyll (BChl) molecules display two major absorption bands: one in the blue/near-UV region (the Soret band, comprising the Bx and By transitions to the second singlet excited state) and one in the red/near-IR region (the Q band, comprising the Qx and Qy transitions to the first singlet excited state). Both transitions involve the $\pi$ electrons of the conjugated chlorin macrocycle. The energy of these transitions is sensitive to the molecular environment: solvent polarisability, nucleophilicity, electrophilicity, excitonic interactions with neighbouring pigments, and axial coordination of the central magnesium atom all shift band positions (*16-18*). This environmental sensitivity is what makes (B)Chls informative biological probes and what makes resonance Raman well-suited to interrogating them in their native protein context.

Resonance Raman spectra of (B)Chls contain approximately 60 bands distributed across three frequency regions, each reporting on a distinct structural aspect of the molecule (Figure 2C). The low-frequency region (50–700 $cm^{-1}$) contains modes involving the central Mg ion. Bands near 210 and 130 $cm^{-1}$ are attributable to Mg–N motions, and a mode near 290–320 $cm^{-1}$ involves coupled displacements of the magnesium and pyrrolic nitrogen atoms (*19, 20*).

The mid-frequency region (900–1620 $cm^{-1}$) contains modes sensitive to macrocycle conformation. In vitro, an empirical relationship was established between macrocycle core size and the frequencies of these bands (*21-23*). In proteins, however, these bands are mutually independent, most likely because they report on the precise local geometry of the macrocycle rather than on an average core size, and are sensitive to distortions below 1 Å (*24*). The strong band near 1550 $cm^{-1}$ reports the axial coordination of the central Mg atom: it appears at 1551–1554 $cm^{-1}$ in the five-coordinated species and downshifts to 1545–1548 $cm^{-1}$ in the six-coordinated species, where a second axial ligand binds above and below the chlorin ring (*21, 22*). A band near 1600 $cm^{-1}$, arising from methine bridge stretching and sensitive to the in-plane or out-of-plane position of the Mg atom, shifts to approximately 1615 $cm^{-1}$ when the macrocycle is strained or the Mg is five-coordinated, and falls near 1600 $cm^{-1}$ when it is six-coordinated (*25, 26*). In BChl a, five-coordinated Mg absorbs near 580 nm while six-coordinated Mg is red-shifted to approximately 610 nm; analogous shifts occur in Chl a and Chl b (*27*). Raman is more reliable than absorption alone for assigning coordination state, as absorption shifts can reflect multiple simultaneous environmental effects.

The high-frequency region (1620–1800 $cm^{-1}$) contains the stretching modes of the conjugated carbonyl and vinyl substituents and is the most directly informative region for studying pigment–protein interactions. The $C13^1$ keto carbonyl stretching mode appears near 1700 $cm^{-1}$ for a non-hydrogen-bonded keto group in a non-polar environment and shifts down by up to 45 $cm^{-1}$ according to the strength of the hydrogen bond in which it is involved (*28*). Smaller downshifts, up to 10 $cm^{-1}$, are induced by polar but non-protic environments (*29*) whereas downshifts as large as 55 $cm^{-1}$ indicate double hydrogen bonding (*20*). The stretching modes of the $C7^1$ formyl of Chl b appear near 1661–1664 $cm^{-1}$ (*30*) and those of the $C3^1$ acetyl of BChl a appear near 1655–1665 $cm^{-1}$ under Soret excitation (*31*); both shift down to approximately 1620 $cm^{-1}$ when hydrogen-bonded. A vinyl C=C stretching band near 1620–1625 $cm^{-1}$ is present in Chl a and 2-vinylbacterio-Chl a, but its weak intensity indicates limited conjugation with the macrocycle, consistent with an out-of-plane orientation imposed by steric interactions with the adjacent methyl substituent (*28, 32*).

Together, these three frequency windows allow resonance Raman to report simultaneously on Mg coordination state, macrocycle distortion, and the hydrogen-bonding environment of individual carbonyl groups. The following sections show how this information is extracted from pigment-protein complexes, using the marker band assignments established here as the interpretive framework.

### 3.2. Pigment–protein interactions in photosynthetic antennas

With the vibrational markers of isolated carotenoids and chlorophylls established, it becomes possible to read pigment-protein interactions directly from resonance Raman spectra of intact complexes. Among these markers, the $C13^1$ keto carbonyl stretching mode of Chl *a* is one of the most informative. Its frequency shifts according to the strength of the hydrogen bond formed between the keto group and the surrounding protein: a non-H-bonded keto in a non-polar environment vibrates near 1700 $cm^{-1}$, and the frequency drops progressively (by up to 45 $cm^{-1}$) as the hydrogen bond becomes stronger. The keto stretching region of an LHC protein therefore contains a direct readout of how each Chl *a* is held within its binding pocket.

A clear illustration is provided by LHCII purified with two different detergents, α- and β-dodecyl-maltoside (αDM and βDM, respectively). The two preparations contain the same protein and the same set of pigments, yet their low-temperature absorption spectra differ slightly: a sub-population of Chls-*a* in βDM-LHCII is red-shifted relative to αDM-LHCII, both in the Soret band (430 to 441.5 nm) and in the Qy band (658 to 675 nm), corresponding to shifts of 605 and 382 $cm^{-1}$, respectively (Figure 3A). Resonance Raman spectra recorded under Soret excitation at 406.7 and 413.1 nm reveal the molecular origin of these shifts (Figure 3B and C). The keto region between 1650 and 1700 $cm^{-1}$ contains three groups of contributions: a low-frequency band at 1659 $cm^{-1}$ from strongly H-bonded keto groups, a broad envelope around 1670 to 1680 $cm^{-1}$ from medium-strength H-bonds, and two bands at 1685 and 1700 $cm^{-1}$ from weakly H-bonded or free keto groups. Comparison of the two preparations shows that βDM-LHCII has more intensity around 1670 $cm^{-1}$ and less intensity at 1685 and 1700 $cm^{-1}$ than αDM-LHCII. The change in detergent therefore brings two Chls-*a*, whose keto groups were free or weakly H-bonded in αDM-LHCII, into a configuration in which they form medium-strength hydrogen bonds with the protein (*33*). The two excitation wavelengths used in this experiment provide complementary views. At 406.7 nm, the resonance condition preferentially enhances the modes of Chls *a* whose keto groups are weakly or non-H-bonded; at 413.1 nm, the modes of strongly H-bonded keto groups are favoured (Figure 3B and C). This pattern is itself a manifestation of the resonance principle: Chls *a* with H-bonded keto groups have their Soret absorption red-shifted relative to Chls *a* with free keto groups, and so each excitation wavelength selectively interrogates a different subpopulation. The same logic applies to the Qy band: the red-shift of 382 $cm^{-1}$ observed in absorption tracks the formation of hydrogen bonds at the $C13^1$ position. Hydrogen bonding to the keto carbonyl thus tunes the energy of both the Soret and Qy transitions of protein-bound Chl-*a*, and the magnitude of the shift scales with the strength of the bond. The contributions from macrocycle distortion can be ruled out as a source of this effect. The bands near 1550 and 1600 $cm^{-1}$, which are sensitive to macrocycle strain and to the coordination state of the central magnesium atom, are unchanged between αDM- and βDM-LHCII in both position and width. The differences observed at higher frequency are therefore not due to a change in macrocycle geometry but reflect specifically the hydrogen-bonding environment of the $C13^1$ keto group. These results illustrate how resonance Raman, through a small number of well-characterised marker bands, isolates a single structural variable, in this case the strength of one hydrogen bond, from the broader pigment-protein interaction landscape, and links it quantitatively to a measurable shift in the absorption spectrum.

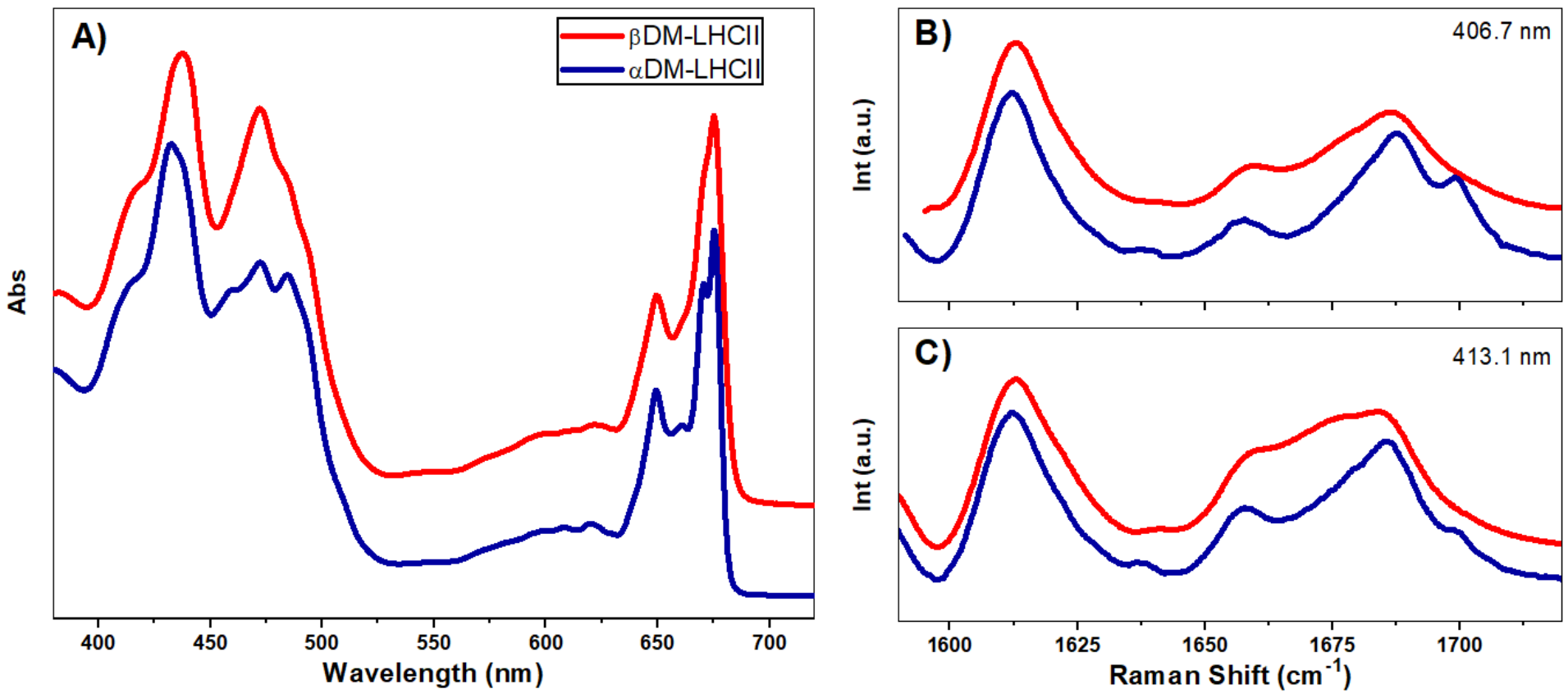

**Figure 3 |** Resonance Raman selectivity reveals hydrogen-bonding differences between two LHCII preparations. **(A)** Low-temperature absorption spectra (4.2 K) of αDM-LHCII (blue) and βDM-LHCII (red). **(B, C)** Resonance Raman spectra of αDM-LHCII (blue) and βDM-LHCII (red) at 77 K in the 1590 to 1720 $cm^{-1}$ region, recorded with excitation at 406.7 nm and 413.1 nm, respectively.

Carotenoid-selective excitation in the same complex addresses a different biological question. The $\nu_4$ region is the primary reporter: an increase in out-of-plane band intensity indicates protein-induced distortion of the carotenoid backbone, and specific $\nu_4$ patterns have been proposed as fingerprints of conformational states associated with non-photochemical quenching. These assignments have been tested under controlled illumination and relaxation protocols in native membranes and reconstituted systems (*34*-*38*). Related complexes provide additional cases where the same interpretive framework applies. In smaller antenna or stress-related assemblies such as high-light-inducible proteins (Hlips), low-temperature RR has been used to compare distortion-sensitive chlorophyll macrocycle bands and carbonyl clusters with the in vitro benchmarks, supporting heterogeneous pigment environments and, in some cases, selective macrocycle distortions not apparent from structural data alone (*39, 40*).

## 4. Femtosecond Stimulated Resonance Raman Spectroscopy

The resonance Raman approach described so far provides a detailed view of the local structure around selected chromophores under equilibrium conditions. However, many biologically relevant processes unfold on ultrafast timescales, ranging from tens of femtoseconds to tens of picoseconds, including excitation-energy transfer, internal conversion, charge separation, vibrational cooling, and the associated structural rearrangements (*41*). Tracking these processes requires techniques capable of resolving vibrational structure with comparable temporal resolution. Transient absorption (TA) spectroscopy is widely used to monitor population dynamics following pulsed photoexcitation, providing access to the temporal evolution of electronic states (*42*). However, TA has limited selectivity: different species can have overlapping electronic signatures, and the same species can show overlapping signatures for different excited states. Time-resolved spontaneous Raman spectroscopy, in principle, overcomes this limitation by directly probing vibrational structure through pump–probe schemes that combine an actinic excitation pulse with a delayed Raman probe. In practice, however, this approach faces significant constraints. The time–bandwidth limit imposed by the Heisenberg uncertainty principle restricts the simultaneous achievement of high temporal and spectral resolution (*43*), hindering the investigation of sub-picosecond dynamics in systems with congested vibrational spectra. The shorter the pulse, the broader the range of energies it carries, and a 100 fs pulse cannot separate Raman bands lying closer than about 150 $cm^{-1}$ (*3*) (Figure 4). In addition, the inherently weak scattering efficiency of spontaneous Raman processes leads to low signal levels when using pulsed excitation. Although resonance enhancement can increase the Raman cross section, it also introduces strong fluorescence backgrounds that can obscure the vibrational signal. These limitations motivate the development of alternative approaches, such as *femtosecond stimulated Raman spectroscopy* (FSRS), which address these challenges.

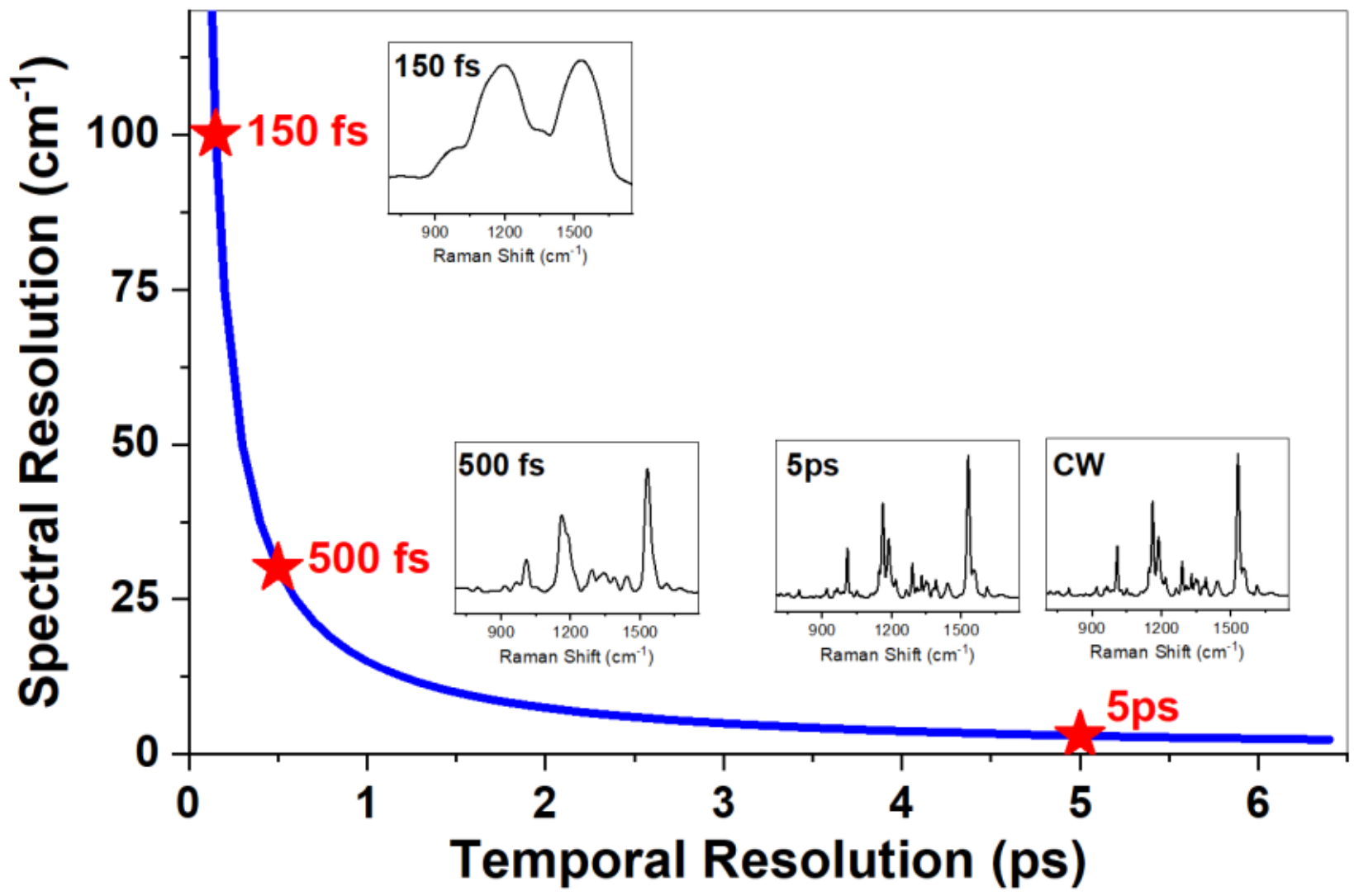

**Figure 4 |** Trade-off between temporal and spectral resolution imposed by the time-bandwidth limit. The theoretical relationship between pulse duration and achievable spectral resolution (blue curve) shows that the two quantities cannot be optimised simultaneously: shorter pulses give better time resolution at the cost of broader spectral features. Red stars mark the three pulses used as example. Insets show simulated Raman spectra under each condition, illustrating the progressive loss of spectral resolution as the pulse duration decreases.

## 4.1. The pulse scheme and the role of resonance

The FSRS experiment involves three laser pulses interacting with the sample (Figure 5A). An ultrashort actinic pump (typically 30 to 150 fs) arrives at time zero and initiates photochemistry in the sample. The vibronic activity that follows decays on the timescales of excited-state photophysics. After a controlled delay Δt, a narrow Raman pump (ca. 5 ps duration) and a broadband femtosecond probe, covering a window of 2000 cm$^{-1}$ around the Raman pump (*44*), are overlapped in time and space at the sample. The simultaneous interaction of the Raman pump and the probe with the sample drives stimulated Raman transitions, which imprint sharp gain features on the transmitted probe spectrum at frequencies shifted from the Raman pump by the vibrational modes of the sample (*44*). The stimulated Raman response is retrieved by taking the logarithm of the ratio of probe spectra recorded with and without the Raman pump (*3*, *44*, *45*). Two useful properties follow from this arrangement. First, the time resolution is set by the cross-correlation of the actinic pump and the femtosecond probe, not by the duration of the narrow Raman pump. Second, the width of the Raman bands does not depend on the probe.

Two things set it: how long the Raman pump lasts, and how long the vibration keeps oscillating in step before it dies away (its dephasing time). With picosecond Raman pumps the second is usually the limiting one, and bands are typically 10 to 15 $cm^{-1}$ wide (*44*). Time and frequency resolution are therefore controlled independently, which is the central technical advantage of FSRS over spontaneous time-resolved Raman (*44*).

The energy-level representation of the same experiment is shown in Figure 5B. The actinic pump drives the sample from the ground state ($S_0$) to the first excited state ($S_1$) and prepares a vibrational wavepacket on $S_1$. After the delay Δt, the Raman pump and the probe together induce a stimulated Raman transition that returns the system to a vibrationally excited level of $S_1$, shifted from the initial population by a Raman frequency. The intermediate level reached by the Raman pump is drawn as Sn in the figure for illustration, and this drawing choice is where the distinction between FSRS and FSRRS becomes visible. If the Raman pump couples to a virtual level (no real electronic state at this energy), all species populated at Δt contribute to the stimulated response with similar Raman cross sections, and the experiment is off-resonant. This is standard FSRS. If the Raman pump couples to a real electronic state of the excited-state species (a higher excited state reached from $S_1$, drawn here as Sn), the resonance condition is satisfied, and the Raman cross section of the species whose transient absorption matches the Raman pump is enhanced by several orders of magnitude. This is *Femtosecond Stimulated Resonance Raman Spectroscopy* (FSRRS) (*46*). The resonance mechanism is the same as in ground-state resonance Raman (Section 2); the difference is that the resonant transition is now a transient absorption of an excited-state species rather than a ground-state absorption.

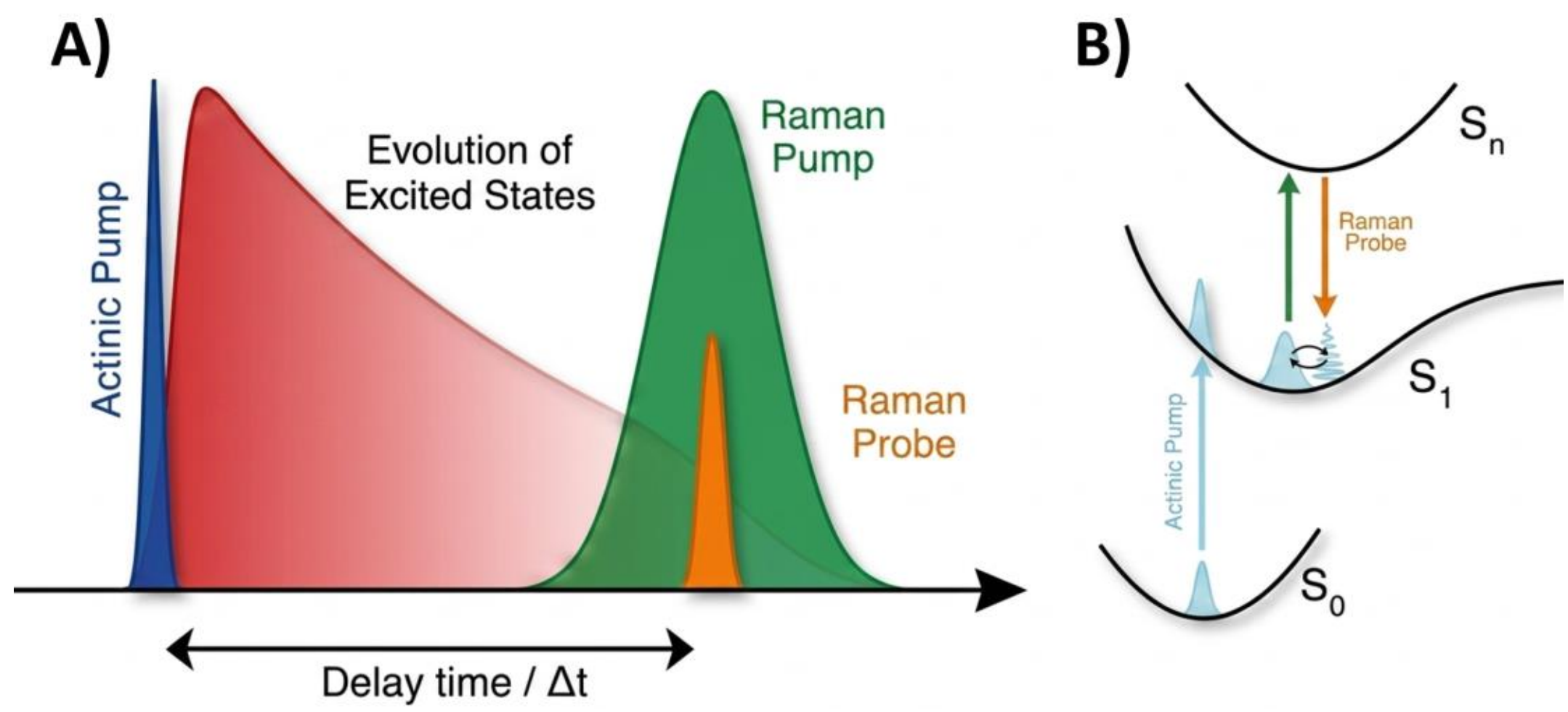

**Figure 5 |** The FSRS pulse sequence and energy-level representation. **(A)** Time-domain view. The actinic pump (blue) initiates photochemistry in the sample. After a controlled delay (Δt), the Raman pump (green, picosecond duration) and the Raman probe (light orange pulse, broadband femtosecond) are overlapped in time and interact simultaneously with the sample. The stimulated Raman response is retrieved from the probe spectrum. **(B)** Energy-level representation. The actinic pump (blue arrow) drives the $S_0 \rightarrow S_1$ transition and prepares a vibrational wavepacket on $S_1$. At delay Δt, the Raman pump (green arrow) and the Raman probe (orange arrow) together induce a stimulated Raman transition that returns the system to a vibrationally excited level of $S_1$, shifted from the initial population by a Raman frequency. The level reached by the Raman pump is shown here as $S_2$ for illustration. If this level coincides with a real electronic state of the sample (here $S_2$), the Raman pump is in resonance with an electronic transition of the excited-state species and the experiment is in the FSRRS regime, with Raman cross sections of the resonant species enhanced by several orders of magnitude. If no real electronic state is present at this energy, the Raman pump couples to a virtual level, the experiment is off-resonant (standard FSRS), and all populated species contribute with similar weight to the measured response.

### 4.2. Tuning the Raman pump across resonance conditions

Whether a given FSRS measurement reports on one species or on many therefore depends on the choice of Raman pump wavelength. In a standard FSRS implementation, the Raman pump is placed in the near-infrared, far from any electronic transition of the sample, and is usually described as off-resonant. This configuration works well when one species dominates the measurement. It is less useful when several excited states with overlapping vibrational frequencies coexist, which is the situation in much of biological photophysics. A near-infrared Raman pump also hides the low-frequency part of the spectrum, because light scattered from the pump itself falls where those bands appear. A visible Raman pump keeps that region clear and reaches bands down to about 100 $cm^{-1}$ (*3*). FSRRS places the Raman pump in resonance with a chosen transient absorption, which makes the Raman pump wavelength an experimental choice that selects which species contributes most strongly to the measured vibrational spectrum (Figure 6). Tunable visible Raman pumps, and the resonance enhancement they provide, have been available since the mid-2000s (*3*). What the approach described here adds is to step the Raman pump across the transient absorption manifold and to treat its wavelength as one more variable in the analysis. In practice, a TA measurement on the same sample guides this choice. The TA spectrum shows where the species of interest absorbs, and where stimulated emission or ground-state bleach would compete with the Raman signal.

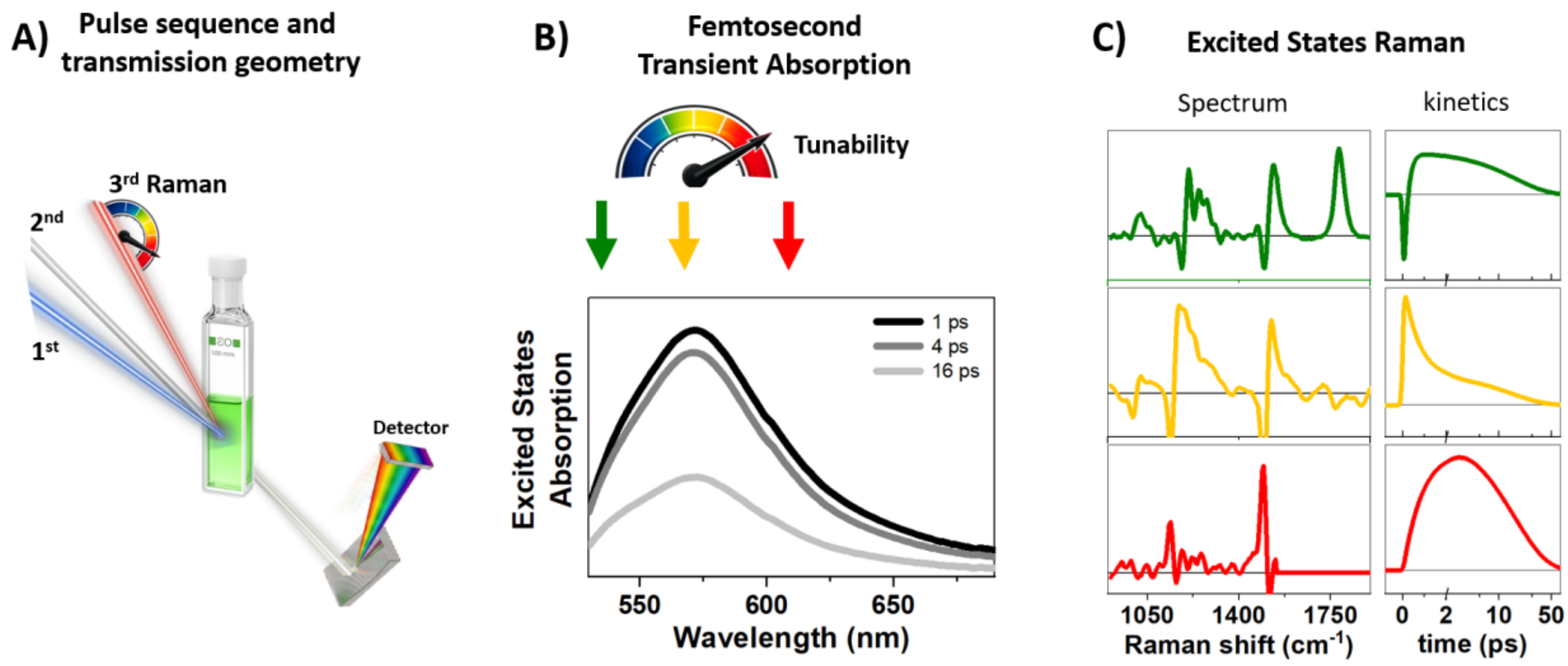

**Figure 6|** Principle of FSRRS and wavelength-selective detection of excited-state contributions. **(A)** Pulse sequence and transmission geometry. An actinic pump (AP; circa 100 fs) initiates photochemistry at t = 0. After a delay τ, a narrow Raman pump (RP; circa 5 ps duration, bands are typically 10 to 15 $cm^{-1}$ wide) and a broadband femtosecond probe are overlapped in the sample. **(B)** RP tuning across the transient absorption manifold. The RP wavelength is placed in resonance with different regions of the manifold; under each condition, the Raman cross section of the species whose transient absorption matches the RP is selectively enhanced. **(C)** FSRRS response at the same pump-probe delay under three RP conditions. Recorded on the same sample with only the RP wavelength changed, the three spectra display different vibrational features, showing that the RP selects which transient contributions dominate the response.

## 4.3. The data obtained and global analysis treatment

Three axes carry information in an FSRRS dataset: the vibrational frequency, the pump-probe delay, and the Raman pump wavelength. To illustrate this principle, Figure 7 shows several FSRRS datasets recorded with the same actinic pump and the same delay range, but with the Raman pump placed at different wavelengths chosen to be in resonance with the absorptions of different excited-state species. Each of the datasets contains the same underlying vibrational features, but the relative intensities of those features differ; different species' vibrational features are preferentially enhanced. The number of intermediate Raman pump (RP) wavelengths can be increased to as many as necessary to disentangle the different features. The RP wavelength is an independent experimental variable. It can be sampled at many points across a transient absorption profile, and the resulting dataset carries four variables: the actinic pump wavelength, the Raman pump wavelength, the vibrational frequency, and the pump-probe delay. How densely this space is sampled determines how cleanly coexisting species can be separated. The cost is experimental: every Raman pump wavelength is effectively a new measurement, with its own baseline, its own photodamage control, and its own acquisition time.

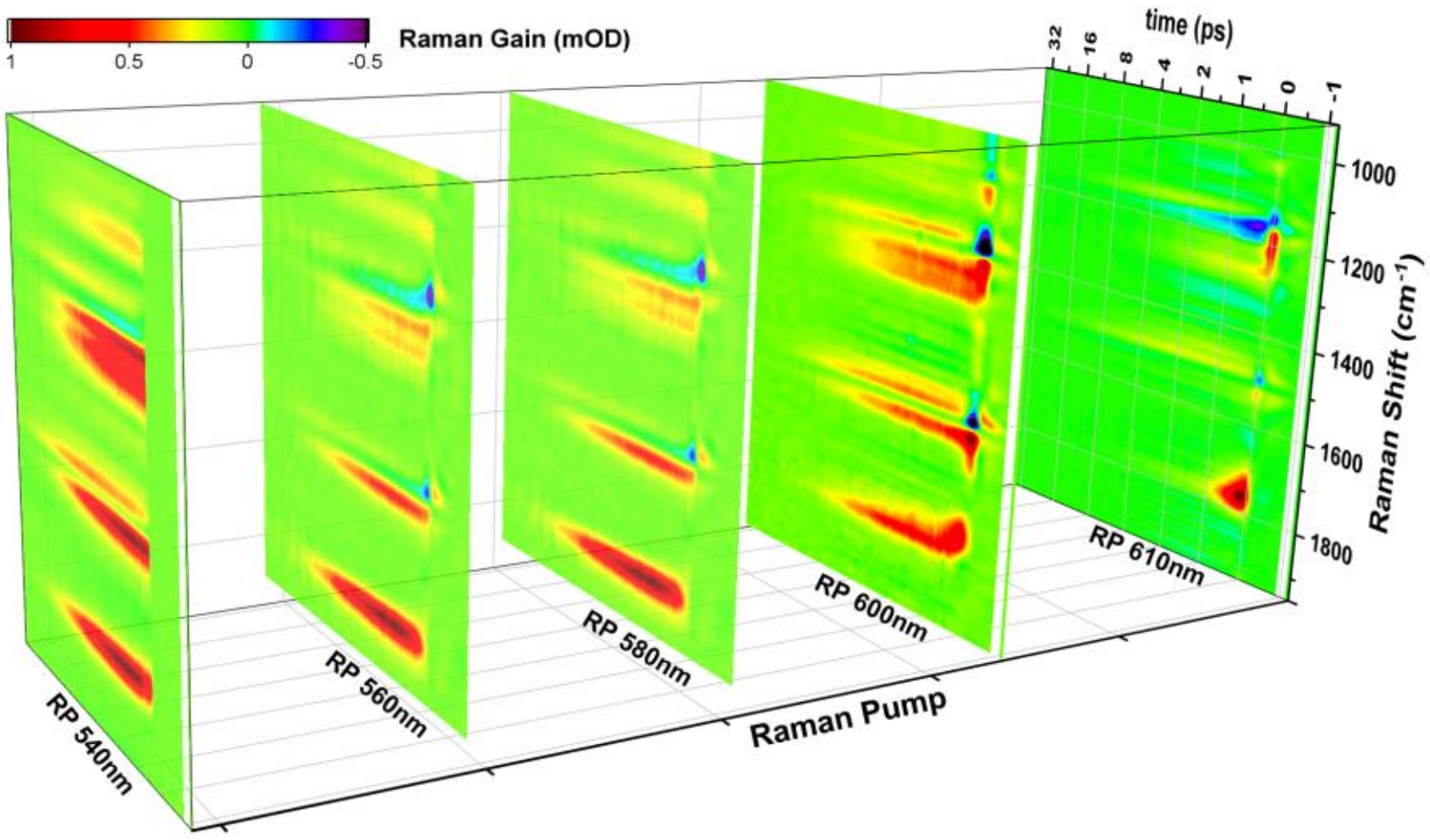

**Figure 7|** Femtosecond stimulated resonance Raman spectroscopy (FSRRS) datasets recorded using the same actinic pump and identical delay range, with the Raman pump (RP) tuned to several wavelengths in resonance with the transient absorption bands of different excited-state species. Although the datasets contain the same underlying vibrational features, the relative intensities vary depending on the RP wavelength, reflecting the preferential resonance enhancement of distinct excited-state populations in each condition.

TA and FSRRS data are commonly analysed with global analysis under simple kinetic models (sequential or parallel) or with target analysis when a more elaborate connectivity is needed (*47, 48*). The output of these procedures is a set of species-associated spectra (SAS) and the time-dependent concentrations of the contributing species under the assumed model. A possible use of the resonance dimension is to perform a single global analysis on multiple FSRRS datasets recorded at different Raman pump wavelengths.

Since the same sample and the same process are probed in each experiment, the same kinetic model is imposed across all datasets, while the relative amplitudes of the contributing species differ because each RP wavelength selectively enhances different species. Multi-dataset global analysis can therefore be performed by fitting the data matrices simultaneously, with each dataset collected at a distinct RP wavelength and the actinic pump wavelength and all other experimental conditions held constant. All resonance conditions probe the same excited-state species, so lifetimes and vibrational band shapes are shared across the fit, and only the amplitudes are allowed to vary between resonance conditions.

Linking the datasets in this way improves the separation of overlapping excited-state species. Components that are weak or short-lived under one resonance condition can be constrained by matrices in which they are better resolved, while the shared spectral fingerprints reduce the number of free parameters and prevent noise-induced spectral variations from being interpreted as genuine changes in the excited-state signatures. Multi-matrix analysis can be performed with available open-access scripts (*49*). The resulting kinetic parameters and SAS are therefore more tightly determined than those obtained from any single dataset (*50*). Comparison with TA lifetimes recorded under the same conditions provides an independent check on the kinetic component of the analysis.

## 5. FSRRS for biologically relevant molecules

Carotenoids are a convenient system to show how FSRRS performs relative to its non-resonant counterpart. They possess the highest resonance Raman cross sections among natural biomolecules, their ground-state vibrational map ($\nu_1$ to $\nu_4$) is well established (*51, 52*), and their excited-state manifold contains several optically dark states whose assignment has been contested for decades (*53-55*). These states share a narrow spectral window in both electronic absorption and vibrational frequency, and their separation by TA alone is rarely unambiguous. Carotenoids therefore provide a practical case for any time-resolved vibrational method that claims to separate coexisting excited species.

Early FSRS studies on isolated β-carotene used picosecond Raman pumps in the near-infrared. This configuration is usually described as off-resonant, although an 800 nm Raman pump overlaps excited-state absorption of the molecule, and the reported features were assigned in part from their resonance enhancement (*44*). It detects the pronounced upshift of the $\nu_1$ C=C stretching band from ca. 1520 $cm^{-1}$ in the ground state to ca. 1780 to 1800 $cm^{-1}$ in the $S_1$ state (*56*), a signature that has since been confirmed across a wide range of carotenoids and solvents using femtosecond configurations (*48, 57, 58*). The same measurements resolved only two excited-state vibrational signatures, assigned to $S_2$ and $S_1$, and were read as support for a two-state relaxation model in long polyenes (*44*). What a single fixed Raman pump wavelength does not deliver is a way to separate species that contribute in the same vibrational window. Because the Raman cross section of all populated species scales similarly under off-resonant conditions, multiple excited states contribute with comparable weight to the measured response, and their decomposition relies entirely on the kinetic model assumed during global or target analysis. For a manifold that contains at least three dark states with overlapping lifetimes and vibrational frequencies, this constraint is insufficient on its own. A second limitation follows from the weak enhancement available at a fixed near-infrared Raman pump: the vibrational signal is weak, only the strongest modes are detectable, and the signal-to-noise ratio for the weaker bands is poor. Modes that fall outside the C=C stretching region, such as the C–C stretches in the 1100 to 1300 $cm^{-1}$ range or the methyl rocking modes near 1000 $cm^{-}$

$^{1}$, are particularly affected, even though these regions carry essential information for distinguishing between species with similar C=C frequencies.

FSRRS lifts both limitations at once. Placing the Raman pump in resonance with a transient absorption band of an excited-state species enhances its vibrational signal by orders of magnitude, while contributions from off-resonant species are comparatively suppressed. By stepping the Raman pump across the full transient absorption manifold, the vibrational fingerprint of each excited state can be probed selectively, and the entire spectral window becomes accessible with sufficient signal-to-noise to extract band positions and kinetics reliably. Figure 8 illustrates this principle on lycopene in tetrahydrofuran. The three FSRRS datasets shown were recorded with the same actinic pump and the same time-delay range, but with the Raman pump tuned to 540, 550, and 610 nm, in resonance with the transient absorption bands of different excited-state species. The same underlying vibrational features are present in all three datasets, but their relative intensities vary with the Raman pump wavelength. A feature strongly enhanced at 540 nm may appear weakly at 610 nm, and vice versa. Features that co-vary across the three datasets are assigned to the same excited state; features that do not must arise from different states. Combined with the multi-dataset global analysis, this pattern of co-variation has been used to resolve four $\nu_1$ contributions in linear carotenoids and to assign them to the $S_1$, vibrationally hot $S_1$, intramolecular charge-transfer, and triplet pair $^{1}$[TT]/S* states (*59, 60*), a level of assignment not attainable from any single FSRRS measurement or from off-resonant FSRS at all.

The same principles apply directly to carotenoids in their native biological context. Pigment-protein complexes contain carotenoids bound in defined orientations and conformations, and the dark states of these carotenoids are central to their biological function: excitation energy transfer to chlorophyll, scavenging of chlorophyll triplets, and quenching of excess excitation during photoprotection. Each of these processes involves a specific subset of the dark state manifold, and a state-selective vibrational probe is needed to follow them. Because FSRRS retains its selectivity once the carotenoid is embedded in a protein, the approach can in principle be transferred to light-harvesting complexes, reaction centres, and stress-related antenna proteins, where it may resolve which dark state participates in which biological pathway (*50*).

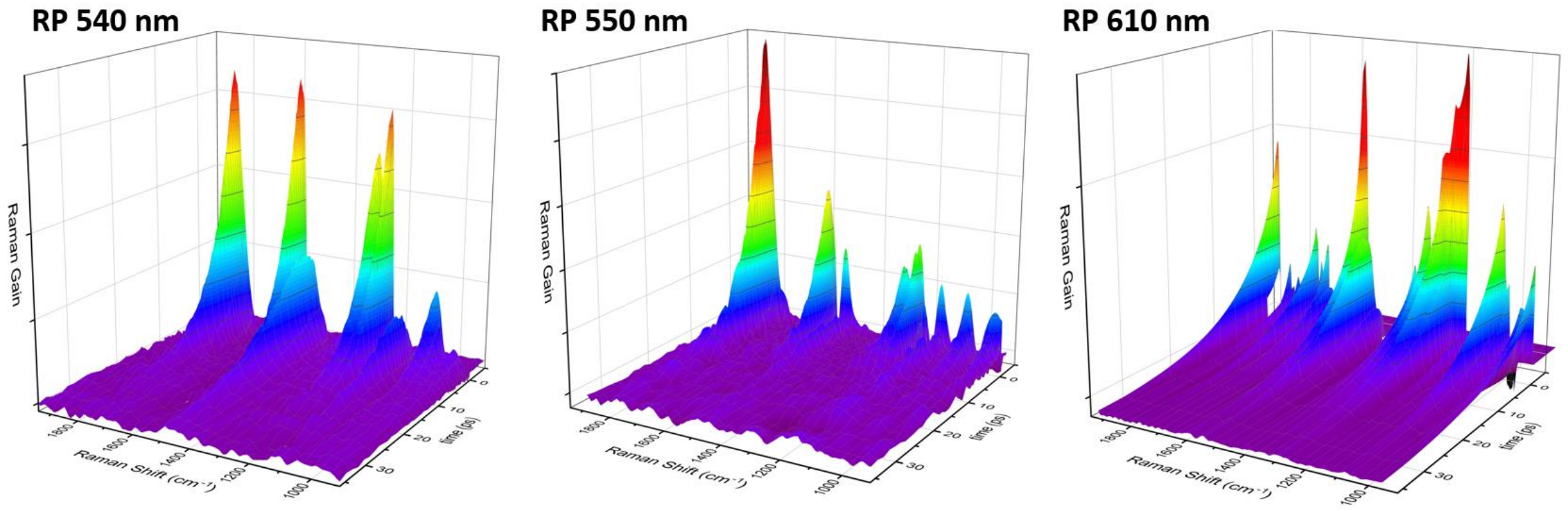

**Figure 8|** Femtosecond stimulated resonance Raman spectroscopy (FSRRS) of lycopene in tetrahydrofuran. The three datasets were recorded on the same sample, with the actinic pump and the delay range held constant and the Raman pump (RP) set to 540, 550 and 610 nm, each in resonance with the transient absorption of a different excited-state species. Band positions are common to the three panels; band amplitudes are not. A mode that is strong at 540 nm can be weak at 610 nm, and it is this pattern of co-variation across RP wavelengths that separates the contributions of distinct excited states.

## 6. Practical considerations for measurements

FSRRS is a delicate but powerful spectroscopic technique capable of extracting structural dynamic features in fragile biological molecules (*50*, *61*). However, there are several experimental constraints that apply to this technique on biological samples in order to obtain sound results. These restrictions fall into four groups: sample preparation, beam alignment and overlap, data acquisition, and data analysis. This section summarises the ones that appear across applications on a regular basis, not as a full protocol but as the points at which measurements most frequently go wrong.

Sample preparation. The optical density at the actinic pump wavelength should be in the 0.5 to 1 range at the path length of the sample cell (typically 0.5 for 1 mm cuvettes or  flow cells), which gives sufficient absorption for a reasonable signal without introducing reabsorption effects on the probe. The optical density at the Raman pump wavelength should be below 0.3, to avoid substantial Raman-pump-induced transient absorption and to limit photodamage. The sample must be continuously refreshed during the measurement: a peristaltic flow cell, or a spinning cuvette are acceptable, provided the refresh time at the interaction volume is shorter than the interval between actinic pump pulses (typically 10 kHz repetition rate). The ground-state absorption spectrum of the sample is recorded before and after each acquisition block, and the two spectra are compared to verify that no cumulative photodamage has occurred.

Beam alignment and overlap. The three pulses must overlap in time and in space in the sample, and the quality of this overlap determines both the time resolution and the signal level. Spatial overlap is typically verified with a pinhole at the sample position and fluorescence imaging of the overlap region. Temporal overlap of the actinic pump and the probe (the pump-probe zero) is determined from the response of a non-resonant sample (typically a solvent or a known dye) and corrected for probe chirp across the detection window. Temporal overlap of the Raman pump and the probe is adjusted to maximise the stimulated Raman signal from a reference compound (cyclohexane and toluene are standard choices). The actinic and Raman pump polarisations are set either to magic angle (54.7° from the probe) when rotational anisotropy effects are to be minimised, or explicitly parallel or perpendicular when anisotropy is part of the study.

Data acquisition. FSRRS datasets at each Raman pump wavelength are acquired with repeated scans of the pump-probe delay, and with on/off modulation of the actinic pump and the Raman pump to separate the stimulated Raman response from background absorption (*43*). The number of scans required depends on the signal level; typical datasets average 500 to 5000 acquisitions per delay point. The delay range is chosen according to the lifetimes of interest, typically from a few hundred femtoseconds before t = 0 to several times the longest expected lifetime. Time zero is redetermined at each Raman pump wavelength, because small changes in beam path during Raman pump tuning can shift the zero by tens of femtoseconds.

Data analysis. The stimulated Raman response is retrieved from the logarithm of the ratio of probe spectra acquired with and without the Raman pump, as described in Section 4. The broad baseline from transient absorption, stimulated emission, and residual scatter is subtracted; polynomial subtraction is the most common approach, although explicit modelling against a separately recorded TA dataset is more robust in cases where the baseline varies substantially with delay. The resulting FSRRS dataset is a function of pump-probe delay and vibrational frequency, and is amenable to global or target analysis. Kinetic components returned by global or target analysis are checked against the lifetimes of the same sample measured by TA. Fitted spectral features obtained from independent FSRRS datasets at different Raman pump wavelengths are then compared: features that arise from the same electronic state appear in multiple datasets with amplitudes that track the expected Raman pump wavelength dependence, and features that do not should be reassigned or investigated further.

Three practical limitations follow from the same features that give FSRRS its selectivity. First, the Raman pump is tuned into the electronic absorption of the sample, so the actinic and Raman pulses share the chromophore spectral window. Photodamage thresholds are therefore lower than in off-resonant FSRS. Second, the stimulated Raman response sits on top of a broad baseline whose subtraction can affect apparent band positions and intensities at the few per cent level; baseline robustness should be verified by repeating the analysis with more than one subtraction strategy. Third, near time zero, cross-phase modulation between the overlapping pulses produces dispersive features that should not be interpreted as conventional Raman lines. The described quantitative analysis is restricted to delays well beyond the instrument response function.

Ultrafast structural events can be addressed by FSRRS, which probes the dynamics of a photoexcited sample by SRS. Combining a broadband femtosecond probe with a narrowband picosecond Raman pump drives a coherent response in selected vibrational modes with femtosecond timing, and without sacrificing spectral resolution. Placing that Raman pump in resonance with a transient absorption raises the Raman cross section of the corresponding species by several orders of magnitude, compared with a pump placed away from any electronic transition. The SRS signal is emitted collinearly with the probe beam, whereas fluorescence is emitted in all directions, so the fluorescent background is efficiently suppressed. Combined with the structural sensitivity of resonance Raman spectroscopy, this approach enables the reconstruction of transient Raman spectra, which in turn are used to determine the underlying relaxation dynamics. Altogether, FSRRS extends vibrational spectroscopy to sub-picosecond phenomena.

## 7. Acknowledgments

This work was supported by France 2030 PEPR LUMA program ULTRAFAST platform grant ANR-22-EXLU-0002. French Infrastructure for Integrated Structural Biology (FRISBI) grant ANR-10-INSB-05 (I2BC Biophysics platform).

## 8. Competing Interests

The authors declare they have no competing financial or non-financial interests.